\documentstyle[preprint,eqsecnum,aps,prb]{revtex}
\renewcommand{\thesection}{\arabic{section}}
\begin{document}
\draft
\title{\Large\bf THE VARIATIONAL THEORY OF THE PERFECT DILATON-SPIN FLUID
                 IN A WEYL--CARTAN SPACE}
\author{O. V. Babourova\thanks{E-mail:baburova.physics@mpgu.msk.su}
        and B. N. Frolov\thanks{E-mail:frolovbn.physics@mpgu.msk.su}
}
\address{Department of Mathematics, Moscow State Pedagogical University,\\
     Krasnoprudnaya 14, Moscow 107140, Russia}
\maketitle
\begin{abstract}
The variational  theory  of  the  perfect  fluid  with intrinsic spin and
 dilatonic charge (dilaton-spin fluid) is developed. The spin tensor
obeys the classical Frenkel condition.  The Lagrangian density of such fluid
is stated, and the equations of motion of the fluid, the Weyssenhoff-type
evolution equation of the spin tensor and the conservation law
of the dilatonic charge are derived. The expressions of the matter currents
of the fluid (the canonical energy-momentum 3-form, the metric stress-energy
4-form and the dilaton-spin momentum 3-form) are obtained.
\end{abstract}
\newpage

\section{Introduction}
\renewcommand{\thesection}{\arabic{section}}
\markright{The variational theory of the perfect dilaton-spin fluid
           in a Weyl--Cartan space}
     The basic  concept  of  the  modern  fundamental  physics  consists  in
proposition that spacetime geometrical structure is compatible with the
properties of matter filling the spacetime. As a result of this fact the
matter dynamics exhibits the constraints on a metric and a connection of the
spacetime manifold. We shall consider the spacetime with the Weyl--Cartan
geometry and  the  perfect  dilaton-spin  fluid  as  matter  that  fills  the
spacetime, generates the spacetime Weyl--Cartan  geometrical  structure  and
interacts with it.
\par
     The perfect  dilaton-spin  fluid is a perfect fluid, every particle of
which is endowed with intrinsic spin and  dilatonic charge. This model of
the fluid, on the one hand, generalizes the Weyssenhoff--Raabe perfect spin
fluid model\cite{Halb}  and,  on  the  other  hand, can  be  considered  as  a
particular case of the  model  of  the  perfect  hypermomentum  fluid  in  a
metric-affine space (see  Ref.\onlinecite{Los:pr} and references therein).
Significance of matter with dilatonic charge is based on  the  fact  that  a
low-energy effective  string  theory is reduced to the theory of interacting
metric and dilatonic field.\cite{GSW} In this connection the  dilatonic
gravity is  one  of  the  attractive  approaches to the modern gravitational
theory.
\par
     The theory of the perfect fluid with intrinsic degrees of freedom
being developed, the additional intrinsic degrees of freedom of  fluid
particles are  described  with  the  help of a material frame formed by four
vectors $\bar{l}_{p}$ ($p = 1,2,3,4$)  (called `directors'),  adjoined
with each element of the fluid. Three of the directors ($p = 1,2,3$) are
space-like and the fourth ($p = 4$) is time-like.  In some  theories  it  is
accepted that  the  time-like director is collinear to the 4-velocity of the
fluid element.  In this case the fields of the directors form the  frame  of
reference comoving  with  fluid  (the rest frame of reference for each fluid
element).
\par
     The spin intrinsic degrees of freedom are  characterized  by  the  spin
tensor. It is well-known due to Frenkel\cite{Fren} that the spin  tensor
of a  particle  is  spacelike  in its nature that is the fact of fundamental
physical meaning.  This leads to the classical  Frenkel  condition  for a
fluid element,  $S_{pq}u^{q}  =  0$,  where  $u^{q}$ is the 4-velocity of the
fluid element  with  respect  to  the  frame  of  reference  formed  by  the
directors and $S_{pq}$  is  the specific (per particle) spin tensor of the
fluid element with respect to the same frame,  which is calculated  by  some
averaging procedure over the all particles contained into the fluid element.
The Frenkel condition in this form  is  one  of  the  main  principle
that underlies  the  Weyssenhoff  spin fluid theory.\cite{Halb} Validity of
this condition defines the range of applicability of the Weyssenhoff theory.
It should be mentioned that the Frenkel condition appears to be a consequence
of the generalized conformal invariance  of  the  Weyssenhoff  perfect  spin
fluid Lagrangian.\cite{Fr}
\par
     We develop the variational theory of the perfect dilaton-spin fluid  in
a Weyl--Cartan  space  $Y_{4}$ using the exterior form language according to
Trautman.\cite{Tr1} This theory  generalizes the variational theory of the
Weyssenhoff--Raabe perfect spin fluid based on accounting the constraints in
the Lagrangian density of the fluid with the help of Lagrange multipliers
(see Ref.\onlinecite{Los:pr} and references therein).

\section{The dynamical variables and constraints}
\markright{The variational theory of the perfect dilaton-spin fluid
           in a Weyl--Cartan space}
\setcounter{equation}{0} \label{sec:var}
      Let us consider a connected 4-dimensional oriented differentiable
manifold ${\cal M}$ equipped with a linear connection $\Gamma$, a metric
$g$ of index 1 and volume 4-form $\eta$,
\begin{equation}
\eta = \frac{1}{4!}\sqrt{- g}\epsilon_{\alpha\beta\sigma\rho}\theta^{\alpha}
\wedge\theta^{\beta}\wedge \theta^{\sigma}\wedge\theta^{\rho}\;, \qquad
g = \mbox{det} \Vert g_{\alpha\beta}\Vert\;. \label{eq:vol}
\end{equation}
Here $\epsilon_{\alpha\beta\sigma\rho}$ are the components of the totaly
antisymmetric Levi-Civita 4-form density ($\epsilon_{1234}=-1$).  We  use  a
local vector frame $\bar{e}_{\beta}$
($\beta=1,2,3,4$) being nonholonomic in general and a 1-form coframe
$\theta^{\alpha}$ with $\bar{e}_{\beta}\rfloor\theta^{\alpha}= \delta^{\alpha}
_{\beta}$ ($\rfloor$ means the interior product). We shall use according to
Trautman\cite{Tr1} 3-form fields $\eta_{\alpha}$ and 2-form fields
$\eta_{\alpha\beta}$,
\begin{eqnarray}
&&\eta_{\alpha} = \bar{e}_{\alpha}\rfloor \eta = *\theta_{\alpha} \; ,
\qquad \eta_{\alpha\beta} = \bar{e}_{\beta}\rfloor \eta_{\alpha} =
*(\theta_{\alpha}\wedge \theta_{\beta}) \; ,  \nonumber \\
&&\theta^{\sigma} \wedge \eta_{\alpha} = \delta^{\sigma}_{\alpha}\eta \; ,
\qquad \theta^{\sigma} \wedge \eta_{\alpha\beta} = -2\delta^{\sigma}_{[\alpha}
\eta_{\beta ]} \; , \label{eq:3}
\end{eqnarray}
where $*$ is the Hodge dual operator.
\par
     A Weyl--Cartan  space $Y_{4}$ is a space with a curvature 2-form ${\cal
R}^{\alpha}\!_{\beta}$, a torsion 2-form ${\cal T}^{\alpha}$ and with the
metric $g$ and the connection $\Gamma$ which obey the constraint,
\begin{equation}
- {\cal D}g_{\alpha\beta} =: {\cal Q}_{\alpha\beta} = \frac{1}{4}
g_{\alpha\beta}{\cal Q}\;, \qquad
{\cal Q}:= g^{\alpha\beta}{\cal Q}_{\alpha\beta} = Q_{\alpha}\theta^{\alpha}
\;, \label{eq:371}
\end{equation}
where ${\cal Q}_{\alpha\beta}$ is a nonmetricity 1-form,  ${\cal Q}$ is a
Weyl 1-form and ${\cal D}:=d + \Gamma\wedge\ldots$ is the covariant exterior
differential with respect to the connection 1-form $\Gamma^{\alpha}\!_{\beta}$.
\par
     Each fluid element possesses a 4-velocity vector $\bar{u}=u^{\alpha}
\bar{e}_{\alpha}$ which is corresponded to a flow 3-form $u$,
$u:=\bar{u}\rfloor\eta=u^{\alpha}\eta_{\alpha}$ and a velocity 1-form $*\! u=
u_{\alpha}\theta^{\alpha}= g(\bar{u},\underline{})$ with
\begin{equation}
*\!u \wedge u = -c^{2}\eta \;, \label{eq:4}
\end{equation}
that means the usual condition $g(\bar{u},\bar{u}) = - c^{2}$, where
$g(\underline{},\underline{})$ is the metric tensor.
\par
      In the exterior form language the material frame of the directors
turns into the coframe of 1-forms $l^{p}$ $(p = 1,2,3,4)$, which have dual
3-forms $l_{q}$, while the constraint,
\begin{equation}
l^{p} \wedge l_{q} = \delta^{p}_{q}\eta\; , \label{eq:1}
\end{equation}
being fulfilled. This constraint means that
$$
l^{p}_{\alpha}l_{q}^{\alpha} = \delta_{q}^{p}\;, \qquad
l^{p}_{\alpha}l_{p}^{\beta} = \delta_{\alpha}^{\beta}\;,
$$
where the component representations were introduced,
$l^{p} = l^{p}_{\alpha}\theta^{\alpha}$, $l_{q} = l_{q}^{\beta}\eta_{\beta}$.
\par
     The perfect dilaton-spin fluid obeys the Frenkel condition,  which  can
be expressed in two forms,
\begin{equation}
S^{p}\!_{q} u^{q} = 0\;, \qquad u_{p}S^{p}\!_{q} = 0\; . \label{eq:110}
\end{equation}
Here $u^{q} = u^{\alpha}l_{\alpha}^{q}$, $u_{p} = u_{\alpha}l^{\alpha}_{p}$.
The Frenkel conditions (\ref{eq:110}) are equivalent to the equality,
\begin{equation}
\Pi^{p}_{r} \Pi^{t}_{q} S^{r}\!_{t} = S^{p}\!_{q}\;, \quad
\Pi^{p}_{r} := \delta^{p}_{r} + \frac{1}{c^{2}} u^{p} u_{r}\;, \label{eq:121}
\end{equation}
where $\Pi^{p}_{r}$ is the projection tensor which separates the subspace
 orthogonal to the fluid velocity. The equality (\ref{eq:121}) means
that the spin tensor $S^{p}\!_{q}$ is spacelike.
\par
     In the exterior form language the Frenkel conditions (\ref{eq:110})
can be written as
\begin{equation}
S^{p}\!_{q} l_{p}\wedge*\!u= 0 \; , \qquad S^{p}\!_{q} l^{q}\wedge u = 0
\; . \label{eq:12}
\end{equation}
\par
     In case  of  the  dilaton-spin fluid the spin dynamical variable of the
Weyssenhoff fluid is generalized and  becomes  the  new  dynamical  variable
named the  dilaton-spin tensor $J^{p}\!_{q}$:
\begin{equation}
J^{p}\!_{q} = S^{p}\!_{q} + \frac{1}{4} \delta^{p}_{q}J\;, \qquad
S^{p}\!_{q}:= J^{[p}\!_{q]}\;, \qquad J := J^{p}\!_{p}\; . \label{eq:11}
\end{equation}
It is important that only the first term of $J^{p}\!_{q}$ (the spin tensor)
obeys to the Frenkel condition. The second term is proportional to the specific
(per particle) dilatonic charge $J$ of the fluid element. The existence of the
dilaton charge is the consequence of the extension of the Poincar\'{e} symmetry
(with the spin tensor as the dynamical invariant) to the  Poincar\'{e}--Weyl
symmetry with the dilaton-spin tensor as the dynamical invariant.
\par
     A fluid element moving,  the fluid particles number conservation law
(the conservation of the baryon number),\cite{MTW} and the entropy
conservation law are fulfilled
\begin{equation}
d(n u) = 0 \; , \qquad d(n s u) = 0 \; , \label{eq:7}
\end{equation}
where $n$ is the fluid particles concentration equal to the number of fluid
particles per a volume unit, and $s$ is the the specific (per particle)
entropy of the fluid, both in the rest frame of reference.
\par
     The measure of intrinsic motion contained in a fluid element is the
quantity $\Omega^{q}\!_{p}$ which generalizes the intrinsic
`angular velocity' of the Weyssenhoff spin fluid theory,\cite{Halb}
\begin{equation}
\Omega^{q}\!_{p} \eta := u \wedge l^{q}_{\alpha}
{\cal D} l^{\alpha}_{p} \; , \label{eq:9}
\end{equation}
where ${\cal D}$ means the exterior covariant differential with  respect  to
the connection 1-form $\Gamma^{\alpha}\!_{\beta}$,
\begin{equation}
{\cal D} l^{\alpha}_{p} = d l^{\alpha}_{p} + \Gamma^{\alpha}\!_{\beta}
l^{\beta}_{p} \; . \label{eq:10}
\end{equation}
\par
     An element of the perfect dilaton-spin fluid possesses the additional
intrinsic `kinetic' energy density 4-form,
\begin{equation}
E =  \frac{1}{2} n J^{p}\!_{q}\Omega^{q}\!_{p}\eta = \frac{1}{2} n
S^{p}\!_{q} u\wedge l^{q}_{\alpha}{\cal D} l^{\alpha}_{p} +
\frac{1}{8} n J u\wedge l^{p}_{\alpha}{\cal D} l^{\alpha}_{p}\; .
\label{eq:8}\end{equation}
\par
     The internal  energy  density of the fluid $\varepsilon$ depends on the
extensive (additive) thermo\-dy\-namic parameters $n$, $s$ (entropy of a fluid
element per particle),  $S^{p}\!_{q}$,  $J$  and  obeys to the first
thermodynamic principle,
\begin{equation}
d\varepsilon(n, s, S^{p}\!_{q}, J) = \frac{\varepsilon + p}{n} dn +
n T ds + \frac{\partial \varepsilon}{\partial S^{p}\!_{q}} dS^{p}\!_{q}
+ \frac{\partial \varepsilon}{\partial J} dJ \; , \label{eq:13}
\end{equation}
where $p$ is the hydrodynamic fluid pressure, the fluid particles number
conservation law (\ref{eq:7}) being taken into account.

\section{The Lagrangian density and the equations \newline
         of motion of the fluid}
\markright{The variational theory of the perfect dilaton-spin fluid
           in a Weyl--Cartan space}
\setcounter{equation}{0}
   The perfect fluid Lagrangian density 4-form of the  perfect  dilaton-spin
fluid should  be  chosen  as  the  remainder  after subtraction the internal
energy density  of  the  fluid  $\varepsilon$  from  the  `kinetic'   energy
(\ref{eq:8}) with regard to the constraints (\ref{eq:4}), (\ref{eq:12}),
(\ref{eq:7}) which should be introduced into  the  Lagrangian
density by means of the Lagrange multipliers $\lambda$, $\chi^{q}$,
$\zeta_{p}$, $\varphi$ and $\tau$, respectively. As a result of
the Sec. \ref{sec:var} the Lagrangian density 4-form has the form,
\begin{eqnarray}
&&{\cal L}_{m} = L_{m} \eta = - \varepsilon (n, s, S^{p}\!_{q}, J) \eta +
\frac{1}{2}n S^{p}\!_{q} u\wedge l^{q}_{\alpha}{\cal D} l^{\alpha}_{p}
+ \frac{1}{8} n J u\wedge l^{p}_{\alpha}{\cal D} l^{\alpha}_{p} \nonumber\\
&&+ n u \wedge d\varphi + n \tau u\wedge ds + n \lambda (*\!u \wedge u +
c^{2} \eta)
+ n \chi^{q} S^{p}\!_{q} l_{p}\wedge *\! u + n\zeta_{p}
S^{p}\!_{q}l^{q}\wedge u\;. \label{eq:20}
\end{eqnarray}
\par
     The fluid motion equations and the evolution equation of  the
dilaton-spin tensor are derived by the variation of (\ref{eq:20})
with respect to the independent variables $n$, $s$, $S^{p}\!_{q}$, $J$,
$u$, $l^{q}$ and  the  Lagrange  multipliers,  the  thermodynamic
principle (\ref{eq:13}) being taken  into  account. We shall consider the
1-form $l^{q}$  as  an  independent  variable  and  the  3-form $l_{p}$ as a
function of $l^{q}$ by means of (\ref{eq:1}). As a result of such
variational machinery  one gets the constraints (\ref{eq:4}),
(\ref{eq:12}), (\ref{eq:7}) and the following variational equations,
\begin{eqnarray}
\delta n : &&\quad  \frac{1}{2}n S^{p}\!_{q} u\wedge l^{q}_{\alpha}
{\cal D} l^{\alpha}_{p} + \frac{1}{8}n J u\wedge l^{p}_{\alpha}{\cal D}
l^{\alpha}_{p} + n u\wedge d\varphi = (\varepsilon + p) \eta
\; ,\label{eq:21}\\
\delta s : &&\quad  T \eta + u \wedge d\tau = 0 \;, \label{eq:22}\\
\delta S^{p}\!_{q} : &&\quad \frac{\partial \varepsilon}
{\partial S^{p}\!_{q}}\eta = \frac{1}{2}n\Omega^{[q}\!_{p]}\eta + n\chi^{[q}
l_{p]}\wedge *\! u + n \zeta_{[p}l^{q]}\wedge u\; ,\label{eq:23}\\
\delta J : &&\quad \frac{\partial \varepsilon}{\partial J}\eta =
\frac{1}{8}n\Omega^{p}\!_{p}\eta \; ,\label{eq:24}\\
\delta u : &&\quad d\varphi + \tau ds - 2\lambda *\! u + \chi^{q}
S_{\beta q}\theta^{\beta} - \zeta_{p}S^{p}\!_{q} l^{q} \nonumber \\
&& + \frac{1}{2}S^{p}\!_{q} l^{q}_{\alpha} {\cal D}l^{\alpha}_{p}
+ \frac{1}{8}J l^{p}_{\alpha} {\cal D}l^{\alpha}_{p} = 0\; , \label{eq:25} \\
\delta l^{q} : &&\quad \frac{1}{2}\dot{S}^{\sigma}\!_{\rho}l_{q}^{\rho}
\eta_{\sigma} + \frac{1}{8} \dot{J} l_{q} - \chi^{r}S^{p}\!_{r} u_q l_p
 - \zeta_{r} S^{r}\!_{q} u = 0\;. \label{eq:26}
\end{eqnarray}
Here the `dot' notation for the tensor object $\Phi$ is introduced,
$\dot{\Phi}^{\alpha}\!_{\beta} := *\!(u\wedge {\cal D}\Phi^{\alpha}\!_
{\beta})$.
\par
     Multiplying the equation (\ref{eq:25}) by $u$ from the left
externally and using (\ref{eq:21}) one derives the expression for the
Lagrange multiplier $\lambda$,
\begin{equation}
2 n c^{2} \lambda = \varepsilon + p \; . \label{eq:28}
\end{equation}
\par
     As a consequence of the equation (\ref{eq:21}) and the constraints
(\ref{eq:4}), (\ref{eq:1}), (\ref{eq:12}), (\ref{eq:7}) one can verify
that the Lagrangian density 4-form (\ref{eq:20}) is proportional to the
hydrodynamic fluid pressure, ${\cal L}_{m} = p \eta$.

\section{The evolution equations of the spin tensor and the dilatonic
      charge}
\markright{The variational theory of the perfect dilaton-spin fluid
           in a Weyl--Cartan space}
\setcounter{equation}{0}
     The variational   equation   (\ref{eq:26})   represents  the  evolution
equation of the directors. Multiplying the equation
(\ref{eq:26}) by $l^{q}_{\beta}\theta^{\alpha}\wedge\dots$ from the left
externally one gets,
\begin{equation}
\frac{1}{2}\dot{S}^{\alpha}\!_{\beta} +\frac{1}{8}\dot{J} \delta^{\alpha}
_{\beta} - \chi^{r}S^{\alpha}\!_{r}u_{\beta} - \zeta_{r}S^{r}\!_{\beta}
u^{\alpha} = 0\; . \label{eq:30}
\end{equation}
\par
     The contraction (\ref{eq:30}) on the indices $\alpha$ and $\beta$
gives with the help of the Frenkel condition (\ref{eq:110}) the dilatonic
charge conservation law,
\begin{equation}
\dot{J} = 0\;. \label{eq:36}
\end{equation}
\par
     Contracting (\ref{eq:30}) with $u_{\alpha}$ and then with $u^{\beta}$
and taking   into   account   (\ref{eq:36})   and   the   Frenkel  condition
(\ref{eq:110}), one gets the expressions for the Lagrange multipliers,
\begin{equation}
\zeta_{r}S^{r}\!_{\beta} = - \frac{1}{2c^{2}} \dot{S}^{\gamma}\!_{\beta}
u_{\gamma}\; , \qquad  \chi^{r}S^{\alpha}\!_{r} = - \frac{1}{2c^{2}}\dot{S}
^{\alpha}\!_{\gamma}u^{\gamma}\; . \label{eq:32}
\end{equation}
As the consequence of (\ref{eq:36}) and (\ref{eq:32})
the equation (\ref{eq:30}) yields the evolution equation of the spin tensor,
\begin{equation}
\dot{S}^{\alpha}\!_{\beta} + \frac{1}{c^{2}} \dot {S}^{\alpha}\!_{\gamma}
u^{\gamma} u_{\beta} + \frac{1}{c^{2}} \dot {S}^{\gamma}\!_{\beta}
u_{\gamma}u^{\alpha} = 0\; . \label{eq:33}
\end{equation}
This equation generalizes the evolution equation of the spin tensor in the
Weyssenhoff fluid theory to a Weyl--Cartan space. With the help of the
projection tensor (\ref{eq:121}) the equations (\ref{eq:33}) and
(\ref{eq:36}) can be represented in the equivalent form,
\begin{equation}
\Pi^{\alpha}_{\sigma}\Pi^{\rho}_{\beta} \dot{J}^{\sigma}\!_{\rho} = 0 \;,
\label{eq:35}\end{equation}
which is the evolution equation of the total dilaton-spin tensor.

\section{The energy-momentum tensor of the perfect \newline
          dilaton-spin fluid}
\markright{The variational theory of the perfect dilaton-spin fluid
           in a Weyl--Cartan space}
\setcounter{equation}{0}
     By means of the variational derivatives of the matter Lagrangian density
(\ref{eq:20}) one can derive the external matter currents which are the
sources of a gravitational field. In case of the perfect dilaton-spin fluid
the matter currents are the canonical energy-momentum 3-form $\Sigma_{\sigma}$,
the metric stress-energy 4-form $\sigma^{\alpha\beta}$, the dilaton-spin
momentum 3-form ${\cal J}^{\alpha}\!_{\beta}$.
\par
     The variational derivative of the Lagrangian density (\ref{eq:20}) with
respect to $\theta^{\sigma}$ yields the canonical energy-momentum
3-form,\cite{Tr1}
\begin{equation}
\Sigma_{\sigma} := \frac{\delta{\cal L}_{m}}{\delta \theta^{\sigma}}
 = - \varepsilon\eta_{\sigma} + 2\lambda n u_{\sigma} u
+ 2 c^{2}\lambda n \eta_{\sigma} - n \chi^{r}S^{q}\!_{r}(g_{\sigma\rho}
l^{\rho}_{q} u + u_{\sigma}l_{q}) + \frac{1}{2}n \dot{S}^{\rho}\!_{\sigma}
\eta_{\rho}\; . \label{eq:350}
\end{equation}
Using the explicit form of the Lagrange multiplier (\ref{eq:28}), the Frenkel
condition (\ref{eq:110}) and the dilatonic charge conservation law
(\ref{eq:36}), one gets,
\begin{eqnarray}
&&\Sigma_{\sigma} =  p \eta_{\sigma} + \frac{1}{c^{2}}(\varepsilon + p)
u_{\sigma} u + \frac{1}{2} n \dot{S}^{\rho}\!_{\sigma}\eta_{\rho}
- n \chi^{r} S^{\rho}\!_{r}(g_{\sigma\rho} u +
l^{q}_{\rho}l_{q}u_{\sigma})\; . \label{eq:351}
\end{eqnarray}
On the basis of the evolution equation of the spin tensor (\ref{eq:33}) and
with the help of (\ref{eq:32}) the expression (\ref{eq:351}) takes the form,
\begin{equation}
\Sigma_{\sigma} = p\eta_{\sigma} + \frac{1}{c^{2}}(\varepsilon + p) u_
{\sigma} u  + \frac{1}{c^{2}}n g_{\alpha [\sigma}\dot{S}^{\alpha}\!_
{\beta]}u^{\beta}u \; . \label{eq:363}
\end{equation}
\par
     After some algebra using the constraints (\ref{eq:371}) and
(\ref{eq:110}) one can get that in a Weyl--Cartan space the expression of the
canonical energy-momentum 3-form is simplified,
\begin{equation}
\Sigma_{\sigma} =  p\eta_{\sigma} + \frac{1}{c^{2}}(\varepsilon + p)
u_{\sigma} u + \frac{1}{c^{2}}n \dot{S}_{\sigma}\!_{\rho} u^{\rho} u
\; . \label{eq:37}
\end{equation}
This expression literally coincides with the expression for the canonical
energy-momentum 3-form of the Weyssenhoff perfect spin fluid in a
Weyl--Cartan space.  It should be mentioned that in case of the  dilaton-spin
fluid  the specific energy density $\varepsilon$ in (\ref{eq:37}) contains
the energy density of the dilatonic interaction of the fluid.
\par
     The metric stress-energy 4-form \cite{Tr1} can be derived in the same way,
\begin{eqnarray}
&& \sigma^{\alpha\beta} := 2\frac{\delta {\cal L}_{m}}{\delta g_{\alpha\beta}}
= T^{\alpha\beta} \eta \; , \nonumber \\
&& T^{\alpha\beta} = p g^{\alpha\beta} + \frac{1}{c^{2}}(\varepsilon + p)
u^{\alpha}u^{\beta} + \frac{1}{c^{2}}n \dot{S}^{(\alpha}\!_{\gamma}
u^{\beta )} u^{\gamma} \; .
\label{eq:353}
\end{eqnarray}
\par
     The dilaton-spin momentum 3-form can be obtainted in the following way,
\begin{equation}
{\cal J}^{\alpha}\!_{\beta} := - \frac{\delta{\cal L}_{m}}{\delta
\Gamma^{\beta}\!_{\alpha}} = \frac{1}{2}n \left (S^{\alpha}\!_{\beta} +
\frac{1}{4} J\delta^{\alpha}_{\beta}\right )u = {\cal S}^{\alpha}\!_{\beta} +
\frac{1}{4}{\cal J}\delta^{\alpha}_{\beta}\; , \label{eq:38}
\end{equation}
where the spin momentum 3-form ${\cal S}^{\alpha}\!_{\beta} = {\cal J}^
{ [\alpha}\!_{\beta ]} = (1/2) n S^{\alpha}\!_{\beta}  u$  and  the  dilaton
current 3-form ${\cal J} = {\cal J}^{\lambda}\!_{\lambda} = (1/2) n J u$ are
introduced.

\section{Conclusions}
\markright{The variational theory of the perfect dilaton-spin fluid
           in a Weyl--Cartan space}
        The variational theory of the perfect fluid with intrinsic
spin and dilatonic  charge  has  been  developed.  The  essential
feature of the constructed variational theory consists in using the  Frenkel
condition for the spin tensor only but not for the total dilaton-spin tensor.
The Lagrangian density of the perfect dilaton-spin fluid has been stated and
the equations of motion of the fluid and the evolution equations of the
spin tensor  and the dilatonic charge have been derived.  The expressions of
the matter currents of the fluid (the canonical energy-momentum 3-form,  the
metric stress-energy 4-form and the dilaton-spin momentum 3-form), which are
the sources of a gravitational field in the Weyl--Cartan space-time,  have
been obtained.  In the following paper these expressions  will  be  used  in
deriving the  generalized  Euler-type hydrodynamic equation of motion of
the perfect dilaton-spin fluid and the equation of motion of a test particle
with spin and dilatonic charge in the Weyl--Cartan geometry background.


\vskip 0.2cm
\begin{thebibliography}{10}
\bibitem{Halb}
E. Halbwachs, {\em Th\`{e}orie relativiste des fluides a spin}
(Gauthier-Villars, Paris, 1960).
\bibitem{Los:pr}
O. V. Babourova and B. N. Frolov, ``Variational theory of perfect
hypermomentum fluid'' (LANL e-archive gr-qc/9612055, 1996).
\bibitem{GSW}
M.B. Green,J.H. Schwarz, E. Witten, {\em Superstring Theory}
(Cambridge Univ. Press, 1987).
\bibitem{Fren}
J. Frenkel, {\em Z. f\"{u}r Phys.} {\bf 37}, 243 (1926).
\bibitem{Fr}
B.N. Frolov, ``Generalized conformal invariance and gauge theory of
gravity'', in: {\em Gravity, Particles and Space-Time} (Eds P. Pronin and G.
Sardanashvily, Word Scientific, Singapore, 1996) pp. 113--144.
\bibitem{Tr1}
A. Trautman, {\em Symp. Math.} {\bf 12}, 139 (1973).
\bibitem{MTW}
C.W. Misner,  K.S. Thorne, J.A. Wheeler, {\em Gravitation} (W.H. Freeman and
Company, San Francisco, 1973), vol. 2.
\end{thebibliography}
\end{document}